\documentclass{iau} 
\usepackage{graphicx}
\usepackage{xcolor}
\usepackage{amsmath}
\jname{Nuclear Activity in Galaxies across Cosmic Time}
\title[The quasar main sequence for cosmology]  %% give here short title %%
{The quasar main sequence and its potential for cosmology}
\author[Marziani et al.]   %% give here short author list %%
{P. Marziani$^1$,
  D. Dultzin$^2$, A. Del Olmo$^3$\footnote{AdO acknowledges financial support from Spanish grants AYA2016-76682-C3-1-P and .SEV-2017-0709.},  M. D'Onofrio$^4$, \\ J. A. de Diego$^2$, G. M. Stirpe$^5$, E. Bon$^6$, N. Bon$^6$, B. Czerny$^7$,\\ J. Perea$^3$, S. Panda$^{7,8}$, M. L. Martinez-Aldama$^{3,7}$,    \and\ C. A. Negrete$^2$}
\affiliation{$^1$INAF, Osservatorio Astronomico di Padova, Italy,  email: {\tt paola.marziani@inaf.it};  $^2$Instituto de Astronom\'\i a, UNAM, Mexico; $^{3}$IAA (CSIC), Granada, Spain; $^{4}$Universit\`a\ di Padova, Italy;   $^5$INAF, OASS, Bologna, Italy; $^6$ Belgrade Observatory, Serbia; $^{7}$Center For Theoretical Physics, Polish Academy of Sciences, Warsaw, Poland; $^{8}$ Nicolaus Copernicus Astronomical Center, Polish Academy of Sciences, Warsaw, Poland.}
\pubyear{2019}
\volume{356}  %% insert here IAU Symposium No.
\editors{M. Povic et al.}
\begin{document}

\maketitle

\begin{abstract}

The main sequence offers a method for the systematization of quasar spectral properties. Extreme FeII emitters (or extreme Population A, xA) are believed to be sources accreting matter at very high rates. They are easily identifiable along the quasar main sequence,  in large spectroscopic surveys over a broad redshift range. The very high accretion rate makes it possible that massive black holes hosted in xA quasars radiate at a stable, extreme luminosity-to-mass ratio. After  reviewing the basic interpretation of   the main sequence, we report on the possibility of identifying  virial broadening estimators from low-ionization line widths, and provide evidence of  the conceptual validity of redshift-independent luminosities based on virial broadening for a known luminosity-to-mass ratio. 
%% add here a maximum of 10 keywords, to be taken form the file <Keywords.txt>
\end{abstract}

%\firstsection % if your document starts with a section,
              % remove some space above using this command.
\section{Introduction}

The concept of the quasar main sequence originated from a Principal Component Analysis of parameters measured on the optical spectra of $\sim$80 PG quasars  (\cite[Boroson  \& Green 1992]{borosongreen92}). The   first eigenvector 1 (E1) computed by the analysis    was found to be  mainly associated with two anti-correlations,  between strength of FeII$\lambda$4570  and prominence of [OIII]$\lambda\lambda 4959,5007$,  and strength of FeII$\lambda$4570 and FWHM of the  HI Balmer line H$\beta$.  In the plane FWHM H$\beta$ vs. prominence of the optical FeII emission (where the FeII prominence is measured by the parameter $R_\mathrm{FeII}$ defined as the intensity ratio of the FeII $\lambda$4570 blend and H$\beta$),   the  occupation of data points representing low redshift quasars takes the form of an elbow-shaped sequence (Fig. 1).   Quasar spectra show a wide range of line widths, profile shapes, $R_\mathrm{FeII}$, line shifts, line intensities which imply differences in line emitting gas dynamics and ionization levels: the main sequence organizes different properties \cite[(Sulentic et al. 2000a)]{sulenticetal00a}.  

Why are these two parameters --   FWHM H$\beta$\ and $R_\mathrm{FeII}$ --   so important? FeII emission extends from UV to the IR and can dominate the thermal balance of the low-ionization part of the broad-line region (BLR, \cite[Marinello et al. 2016]{marinelloetal16}).   FeII emission is self-similar (at least to a first approximation) in quasars but FeII intensity with  respect to H$\beta$ changes  from object to object. The FWHM(H$\beta$) is explained mainly by Doppler broadening, and is related to projection of the velocity field in the low-ionization BLR along the line of sight. There has been a growing consensus that the low-ionization BLR is  predominantly virialized, since the early results of the first major reverberation mapping campaigns \cite[(Peterson \& Wandel 1999)]{petersonwandel99}. Therefore $R_\mathrm{FeII}$ and FWHM H$\beta$\  can be considered  tracers of the physical and dynamical conditions in the low-ionization BLR. \cite[Sulentic et al. (2000)]{sulenticetal00a} introduced the distinction between two Populations.   The FWHM H$\beta$\ $\le$ 4000 km s$^{-1}$ condition selects narrower sources that are  preferentially moderate-to-strong FeII emitters (Population A). Broader sources  (FWHM H$\beta$\ $>$ 4000 km s$^{-1}$, Population B)  tend to have low FeII emission and are believed to be {\em predominantly} sources radiating at lower Eddington ratio ($L/L_\mathrm{Edd}$)  than the quasars of Population A \cite[(Marziani et al. 2018a)]{marzianietal18a}.  

%reviews by 
%\cite[Anders \& Zinner (1993)]{AndersZinner93} and 
%\cite[Ott (1993)]{Ott93}.

%More up-to-date reviews (although only barely including the only recently found pre-
%solar silicates) are by 
%\cite[Zinner (1998)]{Zinner98}, 
%\cite[Hoppe \& Zinner (2000)]{HoppeZinner00}, 
%\cite[Nittler (2003)]{Nittler03}, and 
%\\cite[Zinner (2004)]{Zinner04}. 
%on the basis of a few physical parameters, the most relevant of them being Eddington ratio, and to a second extent, orientation \cite[(Marziani et al. 2001)]{marzianietal01}. The Eddington ratio systematically changes
\vspace{-0.5cm}
\section{The main driver of the quasar main sequence}

%{\underline{\it The Hubble diagram}}
\label{lledd}

Several approaches consistently support a relation between Eddington ratio and $R_\mathrm{FeII}$. For instance, according to the fundamental plane of accreting massive black holes for reverberation-mapped active galactic nuclei (AGNs) -- a relation connecting $R_\mathrm{FeII}$, $L/L_\mathrm{Edd}$\ and a parameter $D$\ dependent on line shape (\cite[Du et al. 2016]{duetal16})  --  $R_\mathrm{FeII}$ $>$1 implies $L/L_\mathrm{Edd} >$ 1.  An independent confirmation is provided by the analysis of the stellar velocity dispersion of the host spheroid, $\sigma_\star$. The $\sigma_\star$ has been used as a proxy for the black hole mass in accordance with   established scaling laws connecting black hole and host spheroid mass \cite[(e.g., Magorrian et al. 1998]{magorrianetal98}). If the AGNs are subdivided in narrow luminosity bins, the $\sigma_\star$ is found to decrease as a function of $R_\mathrm{FeII}$, implying that  $R_\mathrm{FeII}$\ increases as a function of $L/L_\mathrm{Edd}$ \cite[(Sun \& Shen 2015)]{sunshen15}.  The origin of this connection remains unclear at the time of writing. The structure of the emitting regions is probably influenced by the balance between gravitation and radiation forces \cite[(Ferland et al. 2009)]{ferlandetal09}, and by a change in accretion mode with increasing $L/L_\mathrm{Edd}$\  \cite[(Wang et al. 2014)]{wangetal14}. So the correlation may be a secondary effect of structural changes induced on the BLR \cite[(Panda et al. 2019)]{pandaetal19}. Since $R_\mathrm{FeII}$ is also dependent on metallicity \cite[(Panda et al. 2018)]{pandaetal18}, high $L/L_\mathrm{Edd}$\ might be  associated with enriched gas provided to the emitting region by nuclear and circumnuclear star formation \cite[(D'Onofrio \& Marziani 2018)]{donofriomarziani18}. 

%This latter scenario -- no matter how likely -- is still speculative and only partly supported by observational evidence.  

%\begin{figure}[b]
%\begin{center}
%\includegraphics[width=2in]{hu1.eps} \\
%\vspace*{-1.0 cm}
%\includegraphics[width=2in]{hu.eps} 
%\vspace*{-2.0 cm}
% \caption{}
%   \label{fig1}
%\end{center}
%\end{figure}
%\vspace*{-0.25 cm}
\begin{figure}[t!]
\begin{minipage}[t]{0.6\linewidth}
\centering
%\hspace{-2cm}\vspace{-0.5cm}
\includegraphics[width=3.in]{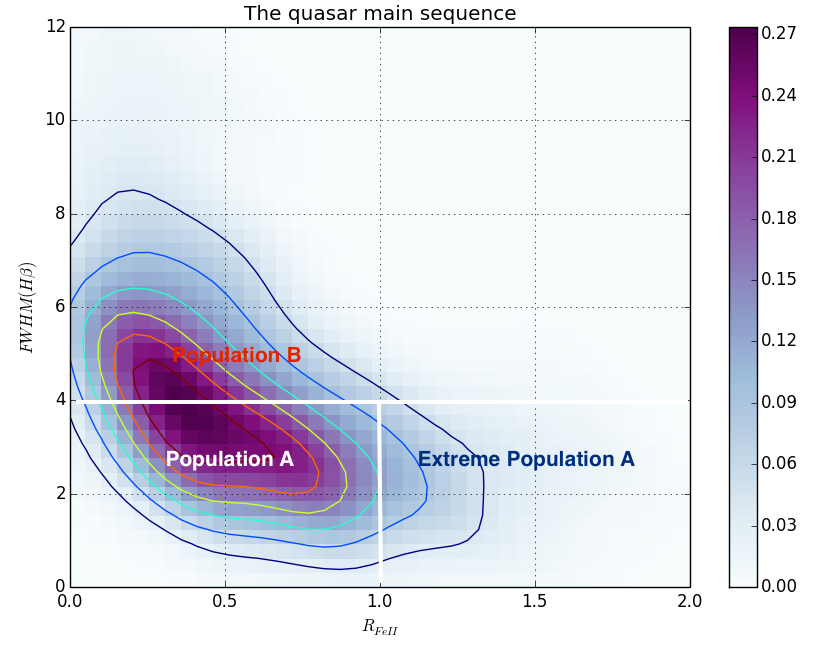}
\end{minipage}
\begin{minipage}[th!]{0.4\linewidth}
\centering
\vspace{-6cm} 
\caption{The optical plane of the E1 MS, FWHM of the broad component of H$\beta$ vs $R_\mathrm{FeII}$. Isodensity curves and shading represent the source occupation from the sample of \cite[Zamfir et al. (2010)]{zamfiretal10}, with $\approx$ 300 quasars. {The labels identify the loci of the main populations: Population A with FWHM\, $\le 4000$ km s$^{-1}$;  Population B with FWHM\, $> 4000$ km s$^{-1}$; extreme Population A,  $R_\mathrm{FeII}> 1$. See text for more details.}}
\label{fig1}
\end{minipage}\vspace{-0.2cm}
\end{figure}

Outflows from mildly-ionized gases producing blue shifted lines   from ionic species of IP $\lesssim 50$\ eV   are ubiquitous in AGNs \cite[(Richards et al. 2011)]{richardsetal11}. The extent and the energetics of the outflows  are not  yet fully appreciated. For example, it has been possible only in recent year to consider high velocity outflows from very hot gas (the so-called ultra-fast outflows,  UFOs, \cite[Tombesi et al. 2011]{tombesietal11}).  However, the prominence of outflows traced by the mildly-ionized gas emitting CIV$\lambda$1549 and [OIII]$\lambda\lambda$4959,5007   increases   along the main sequence  (MS, see Figure 4 of \cite[Sulentic et al. 2000 for CIV]{sulenticetal00}) and reaches a maximum in correspondence of the strongest FeII emitters, at extreme values of $L/L_\mathrm{Edd}$. 

Several interpretations of the MS  (not necessarily in contradiction among themselves) have been proposed.   The occupation of data points can be accounted for by a trend between Eddington ratio and $R_\mathrm{FeII}$, convolved with the effect of orientation on the FWHM of H$\beta$ \cite[(Marziani et al. 2001, Shen \& Ho 2014)]{marzianietal01,shenho14}. This is especially true in the case of a small range of black hole masses. There is a degeneracy between the effect of black hole mass and orientation as both tend to increase the FWHM of H$\beta$\ \cite[(Marziani et al. 2018a)]{marzianietal18a}. \cite[Panda et al. (2019)]{pandaetal19} showed that, for a fixed black hole mass, there is a limit in FWHM beyond which the orientation effects cannot go. Even if limits to  the FWHM range spanned by orientation broadening can be set, the overlap of mass and orientation effects implies that in the 2D representation of the MS make it impossible to retrieve independent values of these parameters (viewing angle, $L/L_\mathrm{Edd}$, black hole mass) for individual quasars. A 3D representation of the MS or additional constraints are necessary. Another important side effect of the  degeneracy between mass and viewing angle is that  a broad range of black hole mass leads to the MS with a wedge-shaped occupation area \cite[(Shen \& Ho 2014)]{shenho14}. 

An alternative view considers that   Population B quasars are more massive  and radiate at  lower values of the $L/L_\mathrm{Edd}$\ than quasars belonging to Population A.  This difference between the two populations  helps define the shape of the main sequence {\em as a sequence} in the optical plane, and  is most likely a consequence of the down-sizing of nuclear activity at low redshift \cite[(Fraix-Burnet et al. 2017)]{fraix-burnetetal17}. 
%More massive black holes radiate at lower Eddington ratios than less massive black holes. 
So it is possible to establish an evolutionary connection from the sources of extreme Population  A to the ones of Population B, where the cosmic arrow of time is provided by the black hole mass which can only grow: the larger the mass the older the source. 
\vspace{-0.5cm}

\section{Highly-accreting quasars}
\subsection{Extreme observational and physical properties}
Extreme Population A (xA) quasars satisfy the condition $R_\mathrm{FeII}$ $>$ 1; they are those $\sim$10\%\ of quasars in low-$z$\ ($\lesssim 1$), optically selected samples with extreme FeII emission. Their prevalence is steeply decreasing  toward higher $R_\mathrm{FeII}$ values: in the range 1.0 $\le $ $R_\mathrm{FeII}$ $< 1.5$\ we find $\approx 7$ \%\ of all quasars; in the range 1.5 $\le $ $R_\mathrm{FeII}$\ $<2,  \approx 3$ \%. Quasars with $R_\mathrm{FeII}$\ $>$ 2 account for less than 1\%\ of the optically-selected quasar population (\cite[Marziani et al. 2013]{marzianietal13}).  xA quasars show distinctive features: their UV  continuum is usually  not reddened, they are often with extremely weak UV emission lines  (i.e., weak-lined quasars with W(CIV)$\lambda$1549 $\le$\ 10\AA\ following \cite[Diamond-Stanic et al. 2009]{diamond-stanicetal09}).  The prominent AlIII$\lambda$1860, and very weak CIII]$\lambda$1909 allow for easy UV selection criteria: if  (1) $R_\mathrm{FeII}  >$ 1.0 is satisfied, then (2) UV AlIII $\lambda$1860\-/\-SiIII]$\lambda$1892 $>$ 0.5 \& SiIII]$\lambda$1892\-/\-CIII]$\lambda$1909 $>$ 1. 

Physical parameters are correspondingly extreme. First of all, their $L/L_\mathrm{Edd}$\ is at the high end of the distribution along the main sequence, with small dispersion \cite[(Marziani \& Sulentic 2014)]{marzianisulentic14}. In other words xA quasars selected according to criteria 1 and 2 are extreme radiators, with maximum radiative output per unit mass close to their Eddington limit. This condition is predicted by accretion disk theory at high (possibly super-Eddington)  accretion rates:  radiative efficiency should be low, and  $L/L_\mathrm{Edd}$\ saturate toward a limiting value \cite[(Sadowski et al. 2014, and references therein)]{sadowskietal14}.   The star formation rate (SFR) in the host galaxy as estimated from radio observations can be up to $\sim$\ a few $10^{3} $M$_\odot $\ yr$^{-1}$\ (\cite[Ganci et al. 2019]{gancietal19}, $z \lesssim 1$). The broad emission line intensity ratios in the UV suggest extremely high values for density ($n \gtrsim 10^{12} -10^{13}$ cm$^{-3}$),  very low  ionization (ionization parameter $\sim 10^{-3} - 10^{-2.5}$, \cite[Negrete et al. 2012]{negreteetal12}), and high metallicity  ($Z \gtrsim$ 20 Z$_\odot$, \cite[Martinez-Aldama et al. 2018]{martinez-aldamaetal18}). 

Broad emission lines in xA sources are produced by gas apparently enriched  by a circumnuclear Starburst. It is tempting to consider that xAs could be the first unobscured stage emerging from obscured stages of AGN evolution. This hypothesis fits an evolutionary sequence  (\cite[Sanders et al. 1988, Dultzin-Hacyan et al. 2003]{sandersetal88,dultzin-hacyanetal03}): merging and strong interaction lead to accumulation of gas in the galaxy central parsecs, and to coeval Starburst and nuclear activity, whereas winds from massive stars and supernov\ae\ provide enriched material.  Feedback effects on the host galaxies  induced by the extreme outflows of xA quasars can be significant if the AGN luminosity is high, as  the outflow thrust and kinetic powers are dependent on the emission line luminosity. This means that xA quasars could likely be a major factor in galactic evolution at moderate-to-high redshift ($z \gtrsim 1$).
It is however important to stress that xA sources  are not necessarily luminous sources --   xAs are the  low-mass (relatively rare) high accretors in the local Universe  as well as the most distant quasars at $z \gtrsim 6$ \, (\cite[Wang et al. 2019]{wangetal19}). 
%the composite spectra of the  highest redshift quasars closely resemble the composite spectra of \cite[Martinez-Aldama et al. (2018)]{martinez-aldamaetal18}.  \subsection{Extreme quasars as {\em Eddington standard candles}}

Almost symmetric H$\beta$ and AlIII$\lambda$1860\ line profiles coexist with  a CIV$\lambda$1549 profile shifted by several thousand km/s, even at the highest luminosity and when radiation forces predominate over gravity, i.e., when $L/L_\mathrm{Edd}$\ is high \cite[(Sulentic et al. 2017, Vietri et al. 2017, Bischetti et al. 2017)]{sulenticetal17,vietrietal17,bischettietal17}.  A virialized subsystem emitting low ionization lines and a subsystem due to winds or outflows constitute two regions that are, at least in part, appearing as kinematically disjoint in the line profiles. %It is remarkable that  H$\beta$\ remains almost symmetric    

In addition, we can count on   xA quasars'  spectral invariance: intensity ratios remain the same; only the line width increases with luminosity. This implies that the radius of the emitting region should rigorously scale as $L^\frac{1}{2}$: if not, the ionization parameter should change with luminosity. No significant spectral change from very high luminosity to low luminosity sources has been detected yet.  Putting together (1) $L/L_\mathrm{Edd} = const.$, (2) $r \propto L^\frac{1}{2}$ and the virial condition (3) $M_\mathrm{BH} \propto r $FWHM$^{2}$, we obtain a relation linking luminosity and line width as $L \propto$\ FWHM$^{4}$\ \cite[(Marziani \& Sulentic 2014)]{marzianisulentic14}. 

Building the Hubble diagram  distance modulus versus redshift  (see e.g., \cite[Risaliti \& Lusso 2015]{risalitilusso15}) from the redshift-independent virial luminosity estimates with xA quasars over the redshift range $0.1 \lesssim z \lesssim 3$, we obtain a distribution consistent with concordance $\Lambda$CDM (Fig. 2). Constraints on the energy density of matter $\Omega_\mathrm{M}$ (0.30 $\pm$ 0.06) are better than  the ones from supernov\ae, because of the $z\sim2$ coverage of the quasar sample.  The significant scatter of individual measurements  ($\sim$ 1.1 - 1.3 mag)  may be associated with (a)  uncertainty of FWHM, which enters with the 4th power in the luminosity relation; (b) orientation that is likely to be the main source of scatter in the classical  scaling relations \cite[(Marziani et al. 2019)]{marzianietal19}; (c) differences in intrinsic properties of the xA quasars i.e., spectral energy distribution, ionizing photon flux, etc. 

\begin{figure}[t!]
\begin{minipage}[t]{0.6\linewidth}
\centering
%\hspace{-2cm}\vspace{-0.5cm}
\includegraphics[width=2.35in]{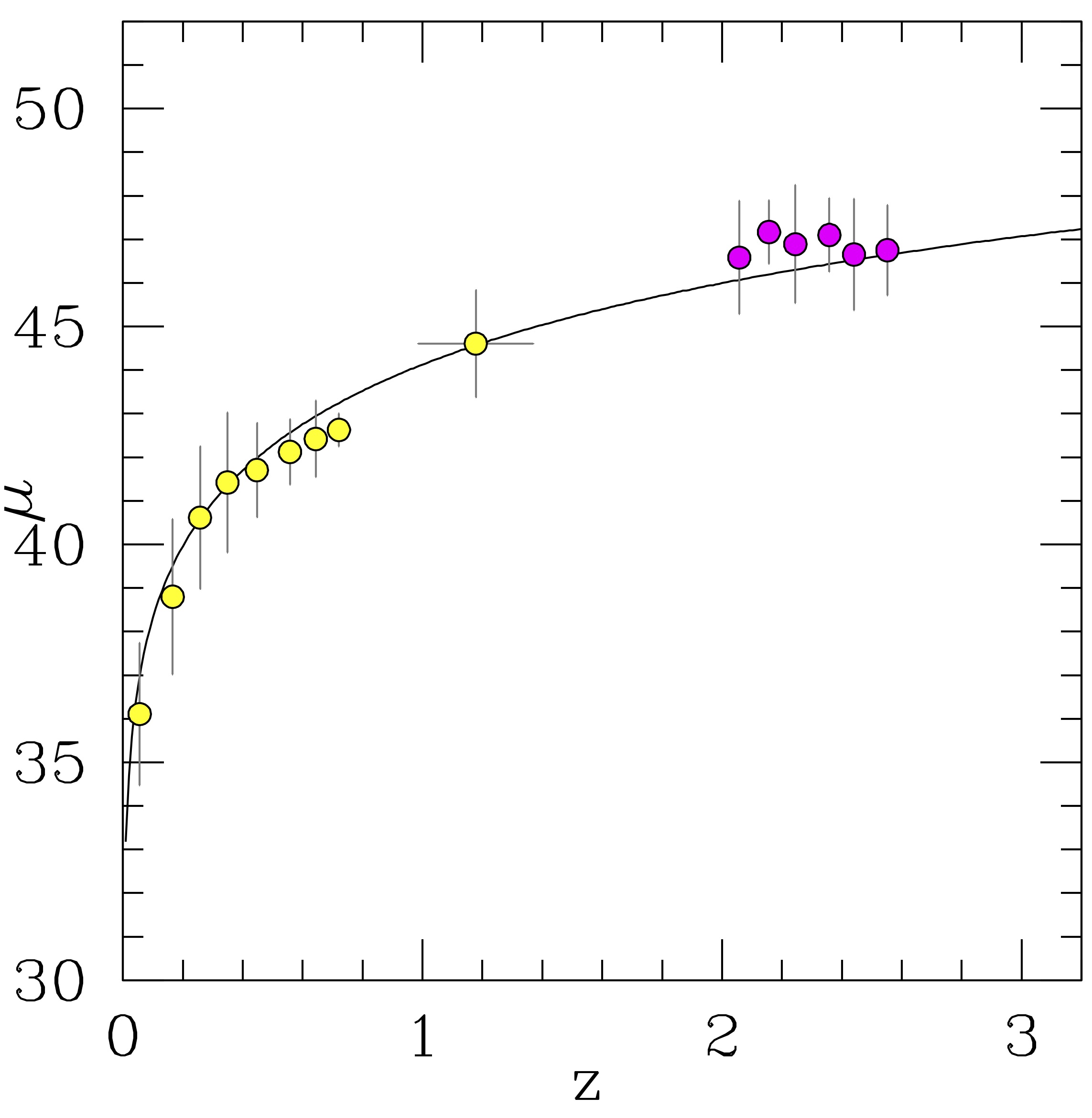}
\end{minipage}
\begin{minipage}[th!]{0.4\linewidth}
\centering
\vspace{-6.5cm} 
\caption{The Hubble diagram distance modulus $\mu$\ vs $z$ for a sample of quasars (\cite[Dultzin et al. 2020]{dultzinetal20}). Individual measurements have been averaged over redshift bins. Error bars are sample standard deviations for $z$ and $\mu$. Yellow data point refer to the use of the H$\beta$ FWHM as a virial broadening estimator, magenta ones are computed from the FWHM of the AlIII$\lambda$1860 doublet individual components. Error bars show sample standard deviations in $\mu$ and $z$\ for each bin.}
\label{fig2}
\end{minipage}
\end{figure}
\vspace{-0.6cm}

\section{Conclusion} 

The MS offer contextualization of quasar observational and physical properties (\cite[Mar\-ziani et al. 2018b]{marzianietal18b}). Several MS trends are ultimately associated  with Eddington ratio which apparently reaches an extreme value in correspondence of the extreme Population A (xA).  xA quasars include the strongest FeII emitters,  satisfying the condition $R_\mathrm{FeII}>1$. They show a relatively high prevalence (10\%)  and are easily recognizable thanks to their peculiar and luminosity-invariant spectral properties. In addition, they are sources that tend to show low intrinsic variability (\cite[Du et al. 2018]{duetal18}), and their low ionization lines are apparently emitted in a virialized BLR (see also Swayamtrupta Panda's contributions in this volume). These properties make xA quasars  suitable as possible {\em Eddington} standard candles, where the invariant properties is not  intrinsic luminosity, but Eddington ratio.
\vfill

%\section{Mock samples}
%\label{mock}

%{\underline{\it Silicon carbide}}. 

\vfill

%\bibitem[Ade et al.(2016)]{adeetal16} Ade, P. A. R., Planck  collaboration 2016, \textit{A\&Ap}, 594, 13

%\bibitem[Faber \& Jackson(1976)]{faberjackson76} Faber, S.~M., \& Jackson, R.~E.\ 1976, \textit{ApJ}, 204, 668 

%\bibitem[Marziani et al.(2001)]{marzianietal01} {Marziani, P.,} Sulentic, J.~W., Zwitter, T., Dultzin-Hacyan, D., \& Calvani, M.\ 2001, \textit{ApJ}, 558, 553 

%\bibitem[Marziani \& Sulentic(2014)]{marzianisulentic14} Marziani, P., \& Sulentic, J.~W.\ 2014, \textit{MNRAS}, 442, 1211 

%\bibitem[Marziani \& Sulentic(2014a)]{marzianisulentic14a} Marziani, P., \& Sulentic, J.~W.\ 2014a, \textit{Adv.  Spa. Res.}, 54, 1331 

%\bibitem[Wang et al.(2013)]{wangetal13} Wang, J.-M., Du, P., Valls-Gabaud, D., Hu, C., \& Netzer, H.\ 2013, \textit{Phys. Rev. Lett.}, 110, 081301 

%\bibitem[Yin, Lee \& Ott (2006)]{Yin_etal06}
%{Yin, Q.-Z., Lee, C.-T. A., \& Ott, U.} 2006, 
 %\textit{ApJ}, 647, 676
%\bibitem[Zinner (1998)]{Zinner98}
%{Zinner, E.} 1998,
 %\textit{Ann. Rev. Earth Planet. Sci.}, 26, 147
%\bibitem[Zinner (2004)]{Zinner04}
%{Zinner, E.} 2004, in: K.K. Turekian, H.D. Holland \& A.M. Davis (eds.), 
 %\textit{Treatise in Geochemistry 1} (Oxford and San Diego: Elsevier), p.\,17%
%\end{thebibliography}

%\bibliographystyle{mnras}
%\bibliography{biblioletter2a}
\end{document}